\documentclass[aps,prl,fleqn,twocolumn,superscriptaddress,floatfix]{revtex4}

\usepackage{epsfig}
\usepackage{amsmath,amssymb,amsfonts}

\begin{document}

\title{Correlated emission of hadrons from recombination of 
       correlated partons}

\author{R.~J.~Fries}
\affiliation{School of Physics and Astronomy, University of Minnesota,
             Minneapolis, MN 55455}

\author{S.~A.~Bass}
\affiliation{Department of Physics, Duke University, Durham, NC 27708}
\affiliation{RIKEN BNL Research Center, Brookhaven National Laboratory, 
             Upton, NY 11973}

\author{B.~M\"uller}
\affiliation{Department of Physics, Duke University, Durham, NC 27708}

\date{\today}

\begin{abstract}

We discuss different sources of hadron correlations in relativistic 
heavy ion collisions. We show that correlations among partons in a 
quasi-thermal medium can lead to the correlated emission of hadrons by 
quark recombination and argue that this mechanism offers a plausible 
explanation for the dihadron correlations in the few GeV/$c$ momentum 
range observed in Au+Au collisions at RHIC. 

\end{abstract}

\preprint{NUC-MINN-04/5-T}

\maketitle

The recombination of thermalized quarks has recently been proposed as 
the dominant mechanism for the production of hadrons with transverse 
momenta of a few GeV/c in Au+Au collisions at the Relativistic Heavy 
Ion Collider (RHIC) \cite{Vo02,Fr03a,Fr04,Gr03a,Hw03a}.
While the concept of recombination of deconfined quarks is not new 
\cite{Gu83}, the RHIC data have provided compelling evidence for the 
presence of this hadronization mechanism. Quark recombination 
explains the enhancement of baryon emission, compared with meson emission, 
in the range of intermediate transverse momenta (roughly from 2 to 5 
GeV/$c$), and it provides naturally for the observed hadron
species dependence of the
elliptic flow in the same momentum 
region in terms of a universal elliptic flow curve for the constituent 
quarks \cite{RHIC_data}.

However, the model of quark recombination from a collectively flowing, 
deconfined thermal quark plasma appears to be at odds with the  
observation of ``jet-like'' correlations of hadrons observed in the 
same transverse momentum range of 2 to 5 GeV/c \cite{STAR:03corr,Sick:04}.
Triggering on a hadron, e.g., with transverse momentum 2.5 GeV/$c < p_T <$ 4 
GeV/$c$, the data shows an enhancement of hadron emission in a narrow angular 
cone around the direction of the trigger hadron in a momentum window below 
2.5 GeV/$c$. Can such correlations be reconciled with the claim 
that hadrons in this momentum range are mostly created by recombination 
of quarks?

Obviously, the observation is incompatible with any model which assumes 
that no correlations exist among the quarks before recombination. Such
correlations require deviations from a global thermal equilibrium in the 
quark phase. One mechanism is already well established: a strong, 
but anisotropic and locally varying, collective flow produces 
correlations among hadrons after recombination. Indeed, the hadronic 
elliptic flow correlation is known to be larger than the elliptic flow
of the quarks before recombination \cite{Vo02,Fr03a}, because 
the parameter $v_2$ characterizing the magnitude of elliptic flow 
for a hadron is proportional to the number of its constituent quarks.

One would generally expect that correlations among the quarks, when
present before their recombination into hadrons, will be amplified by the
hadronization process. 
Here, we argue that a certain amount of two-body
correlations among the partons of a quark-gluon plasma produced in a 
relativistic heavy-ion collision is to be expected. Energetic partons 
produced in such a collision lose a significant amount of energy 
through collisions with thermal partons on their way out of the dense
medium, but do not completely thermalize before they form hadrons. The
dissipated energy and momentum are absorbed by the surrounding medium, 
increasing its temperature slightly and 
setting it into motion in the direction of the energetic parton. This
``wake effect'' produces correlations among medium partons (S) and the 
originally energetic parton (H). If the parton loses enough energy to 
become indistinguishable from the thermal medium, all that remains is 
a narrowly directed, ``jetty'' flow pattern within the thermal medium. 
Such soft 
partons can either recombine with each other upon 
hadronization (denoted by SS for a meson), or a soft parton can recombine
with a hard jet parton (SH) \cite{Gr03a,HwaYa:03}, or a jet parton can 
fragment outside the medium (F) to form a hadron.

\begin{figure}
  \epsfig{file=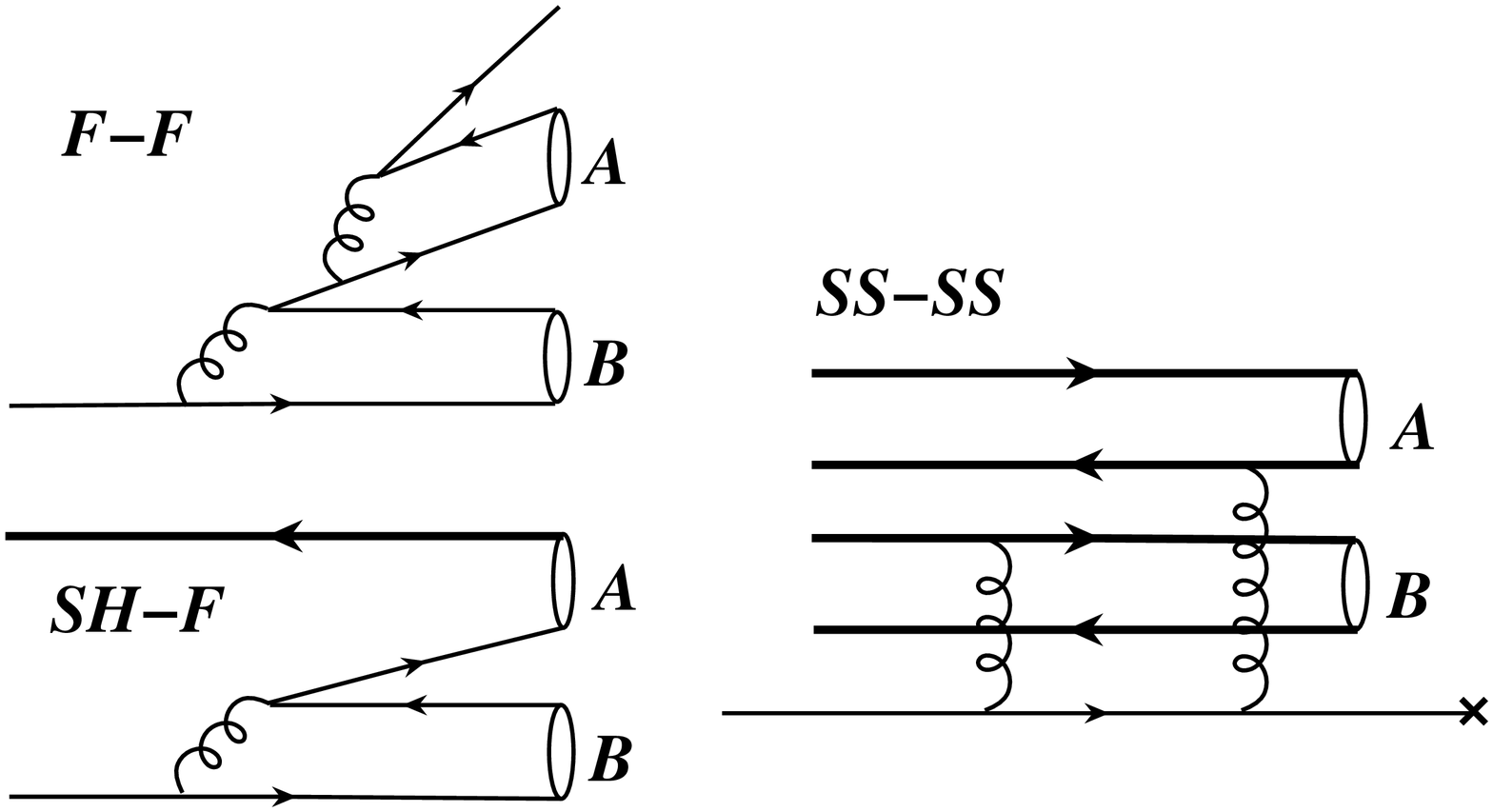,width=6.0cm}
  \caption{\label{fig:processes} Schematic pictures of the processes 
  F-F, F-SH and SS-SS for producing two mesons $A$, $B$. Thick lines are
  medium quarks. Possible correlations are indicated by curly lines.}
\end{figure}

Recently, the STAR collaboration presented experimental evidence that 
jet cones are not isolated from the surrounding medium, but that jets 
and the medium strongly influence each other \cite{STAR-TT}.
They found that hadrons correlated with the jet participate in the 
longitudinal expansion of the medium.
We will not attempt here to give a quantitative description of how the
correlations in the parton phase arise. Rather we will show how such 
correlations between partons can translate into hadron correlations in a 
hadronization scenario with recombination and fragmentation. We will 
focus on dimeson production by three processes: the recombination of 
four thermal partons into two mesons (SS-SS), jet correlations by double 
fragmentation (F-F) and fragmentation accompanied by a soft-hard 
recombination between a jet fragment and a thermal parton (F-SH).
These processes are schematically depicted in Fig.~\ref{fig:processes}. 
Obviously, there are three additional possibilities for producing two 
mesons in this recombination--fragmentation picture, SH-SH, SH-SS and 
F-SS, and even more if  baryons are involved. 
However they do not involve any essentially new aspects and 
we will not consider them further here.

We start by discussing recombination from a thermal, but correlated ensemble
of quarks (SS-SS), following the formalism described byr Fries {\it et al.} 
\cite{Fr03a}. The creation of a pair of mesons $A$, $B$ 
from four partons can be expressed as 
a convolution of the meson Wigner functions $\Phi_A, \Phi_B$ and the 
Wigner function $W_{1234}$ of the four partons, integrated over the 
hadronization hypersurface $\Sigma$:
\begin{equation}
  \frac{E_A E_B \, d^6 N_{AB}}{d^3 {P}_A \, d^3 {P}_B} = C_{AB}
  \int_\Sigma d\sigma_A d\sigma_B \Phi_A \otimes \Phi_B \otimes W_{1234}, 
  \label{eq:1}
\end{equation}
where $C_{AB}=C_A C_B$ is a degeneracy factor. The generalization to the 
case where $A$, $B$, or both are baryons is straightforward. In the 
absence of correlations the $n$-parton Wigner function $W_{1\ldots n}$ 
is approximated by a product of classical one-parton phase space 
distributions $w_i$. This approximation yields satisfactory results for 
inclusive hadron spectra. However, it should not come as a surprise 
that it is not sufficient to describe hadron correlations. For this 
reason, we go one step further and include two-parton correlations
in the form
\begin{equation}
 W_{1234} \approx w_1 w_2 w_3 w_4 \big( 1 + \sum_{i<j}C_{ij} \big).
\end{equation}
Here $C_{ij}$ is the correlation function between partons $i$ and $j$. We 
shall not make special assumptions about the origin of the correlations 
among partons, except that they do not rapidly vary in momentum, can be
localized around a certain direction and are confined to a subvolume 
$V_c$ of the fireball. This is compatible with correlations being induced 
by the interaction of an energetic parton with the thermalized medium, 
as discussed above as a likely source for partonic correlations.

As an example we are going to evaluate the azimuthal correlations 
between mesons emitted around midrapidity in Au+Au collisions at RHIC. 
We assume that momentum correlations of partons are restricted to a 
cone in rapidity $y$ and azimuthal angle $\phi$. 
Hence we choose the correlation functions to be of the form
\begin{multline}
  C_{ij}  =  c_0 \, S_0 \, f_0 
    e^{-(\phi_i-\phi_j)^2/(2 \phi_0^2)} \,
    e^{-(y_i-y_j)^2/(2y_0^2)} \\    
   + c_\pi \, S_{\pi} \,f_{\pi}
     e^{-(\phi_i-\phi_j+\pi)^2/(2 \phi_\pi^2)} \,
    e^{-(y_i-y_j)^2/(2y_\pi^2)} .
\end{multline}
Here $\phi_{0,\pi}$ and $y_{0,\pi}$ are the widths of the Gaussians in 
azimuth and rapidity, respectively. The two terms of the sum correspond 
to correlations initiated by an energetic parton ($\phi=0$) and its recoil 
partner ($\phi = \pi$).  $c_0$ and $c_\pi$ give the strength of the near 
side and far side correlations, while the functions 
$f_{0,\pi}(p_{Ti},p_{Tj})$ describe the transverse momentum dependence of 
the correlations. The functions $S_{0,\pi}(\sigma_i,\sigma_j)$ parametrize 
the spatial localization of the parton correlations on the hypersurface 
$\Sigma$. As indicated above, we assume that $S_{0,\pi} = 1$, if 
$\sigma_i,\sigma_j \in V_c$ and $S_{0,\pi}=0$ otherwise. 

In this letter we restrict the discussion to near side correlations 
($c_\pi = 0 $). We also refrain from exploring the $P_T$ dependence, 
because of the present lack of data, which would allow us to constrain
the function $f_0(p_{Ti},p_{Tj})$. Since we work with fixed $P_T$ 
windows here, for which experimental data exist, we shall simply set 
$f_0 \equiv 1$, absorbing all numerical factors
into the parameter $c_0$. We also assume that $c_0 \ll 1$, allowing us 
to neglect terms of higher power in the correlations, such as $c_0^2$, 
$c_0 v_2$ or $v_2^2$.

Following \cite{Fr03a} we integrate over the spatial coordinates (assuming
$V_{\text{hadron}} \ll V_c \ll V_{\Sigma}$) and the quark momenta 
transverse to the hadron momentum.
Note that, unlike for single 
inclusive hadron spectra, the width of the hadron wave function in the 
azimuthal direction could, in principle, interfere with the correlation 
width $\phi_0$. For simplicity, we neglect such effects here. We also use 
the narrow wave function approximation (see \cite{Fr03a}), 
which was shown to provide a good description of the measured spectra and 
elliptic flow. For the thermal parton distributions $w_i$ we use Boltzmann 
distributions with temperature $T$, radial flow rapidity $\eta_T$ and a boost 
invariant hadronization hypersurface $\Sigma$ at fixed proper time $\tau$ 
\cite{Fr03a}. 

The dimeson spectrum for the SS-SS process is 
\begin{multline}
  \frac{d^6 N_{AB}}{\prod\limits_{A,B} P_{Ti} d P_{Ti} d\phi_i dy_i}  
    = \left( 1 + 2 \hat c_0  + 4 \hat c_0  e^{-(\Delta\phi)^2/(2 \phi_0^2) }
   \right)   
   \\ \times
  \prod_{A,B} h_i(P_{Ti}) \big( 1+ 2 v_{2i}(P_{Ti}) \cos(2 \phi_i) \big) ,  
  \label{eq:2mes}
\end{multline}
where we introduced the abbreviation
\begin{multline}
  h_i(P_T) = C_i \frac{\tau A_T}{(2\pi)^3} M_T 
  I_0 \left(\frac{P_T \sinh \eta_T}{T}\right)\\ \times 
  K_1 \left(\frac{(m_1^T+m_2^T) \cosh\eta_T }{T}\right)
\end{multline} 
with $m^T_j=\sqrt{m_j^2 + P_T^2/4}$. 
$m_j$ are the quark masses and $M_T$ is the transverse mass of the meson.
This quantity is related to the single inclusive meson spectrum at 
midrapidity which, including elliptic flow and two-parton correlations, 
is given by \cite{Fr03a}
$h_i (P_T) (1+\hat c_0)( 1 + 2 v_{2} \cos( 2\phi))$.
The effect of the space-time correlation volume $V_c$ has been approximated
in (\ref{eq:2mes}) by a rescaling of the normalization constant
$\hat c_0 = c_0 V_c/(\tau A_T)$. 

We note that the amplification factor $Q=4$ in front of the Gaussian 
term counts the number of possible correlations of a quark in meson $A$
with a quark in meson $B$. Likewise the term $2\hat c_0$ accounts for
the correlations of quarks inside the same meson. It is then easy to 
see that the amplification factor is $Q=6$ for a meson-baryon pair and 
$Q=9$ for a baryon-baryon pair. This result confirms our expectation 
that correlations within the partonic medium are amplified in the
recombination process, similar to the amplification of the elliptic
flow $v_2$ by the number of valence quarks.

The experiments at RHIC measure the associated particle yield per trigger 
hadron $A$. After subtracting the uncorrelated background and using the
notation $\Delta\phi = |\phi_A - \phi_B|$, the relevant observable is
defined as
\begin{equation}
  Y_{AB}(\Delta \phi) = N_A^{-1} \left(
  {\frac{dN_{AB}}{d(\Delta \phi)} - \frac {d (N_A N_B)}{d(\Delta \phi)}}
\right).
\end{equation}
The particle yields are integrated over the kinematic windows of the
hadrons $A$ and $B$ with exception of the relative angle $\Delta\phi$.
The $P_T$ spectra of associated particles have recently been studied in
\cite{HwaYa:04cor} in a recombination picture.
Neglecting quadratic terms of correlation coefficients, the background 
subtraction cancels the term proportional to $1+2\hat c_0$ in 
Eq.\ (\ref{eq:2mes}) and leads to the result
\begin{equation}
  N_A Y_{AB}(\Delta \phi) 
  = Q \hat c_0 e^{-(\Delta\phi)/(2\phi_0^2) } \,
    N_A N_B/(2\pi) ,
\end{equation}
were $N_i = 2\pi \int dy_i dP_{Ti}P_{Ti} h_i (P_{Ti})$ is the total 
particle number for species $i$ in the kinematic window. The effect 
of a possible correlation in rapidity --- not discussed here --- 
can be absorbed into the normalization $\hat c_0$. The uncorrelated 
background up to first order in the coefficients $\hat c_0$,
$v_2$ is given by $(2\pi)^{-1}N_A N_B [1 + 2\hat c_0 + 2 \bar v_{2A} 
\bar v_{2B} \cos(2 \Delta\phi)]$,
where $\bar v_{2i}$ is the average elliptic flow of hadron species $i$ 
in the kinematic window.

Dihadron production through fragmentation from a jet (F-F) is 
described by dihadron fragmentation functions. These have recently been 
discussed by Majumder and Wang \cite{MajWa:04}. 
We assume that they can be factorized into single hadron fragmentation 
functions $D_{a/h}$ with an appropriate scaling of the momentum variable. 
Since the formalism is strictly collinear, we introduce a Gaussian smearing 
in relative azimuthal angle and rapidity of the two hadron momenta. 
Integrating over rapidities we obtain for hadrons emitted around midrapidity
\begin{multline}
  \frac{d (N_A Y_{AB}(\Delta\Phi))}{dP_{TA} dP_{TB}}  = 
  2\pi I (2\pi\phi_0^2)^{-1/2} e^{-(\Delta\phi)^2/(2 \phi_0^2)} \\ \times
  \sum_{a}  \int_{z_0}^{z_1} \frac{d z_A}{z_A(1-z_A)}  
  g_a\left( \frac{P_{TA}}{z_A} +\Delta E \right)
  \\  \times  D_{a/A}(z_A) 
  D_{a/B}\left( \frac{z_A P_{TB}}{(1-z_a) P_{TA}} \right)    
\end{multline}
Here $g_a(p) = E_p d^3N_a/d^3 p$ is the invariant spectrum of the 
fragmenting parton $a$. The factor $I$
contains the integration over the correlated hadron rapidities $y_A$ 
and $y_B$ in their respective windows around midrapidity, assuming that 
the parton spectra $g_a$ are slowly varying. Of course, 
the width $\phi_0$ for this mechanism does not generally 
coincide with that introduced for correlations among the thermal
partons. The kinematic limits are $z_0 = 2 P_{TA}/\sqrt{s}$ and 
$z_1 = P_{TA} /(P_{TA} + P_{TB})$. $\Delta E$ denotes the average 
energy loss of parton $a$ before fragmentation.

For the F-SH process we adopt the following simple model for dimeson 
production. Suppose a hard parton $a$ with initial momentum $p_a$ dresses 
itself with a pair $b \bar b$ (where $b$ is any flavor) and that the pair
$a \bar b$ hadronizes into a meson $A$. Instead of further fragmentation
the remaining parton $b$ can pick up a soft parton $c$ from the medium and 
recombine into another meson $B$. It is clear that this is only the simplest 
possible realization of F-SH. We assume that the production of $A$ can be 
described by a fragmentation function $D_{a/A}$ and we approximate the
correlated emission by a product ansatz, where one of the parton 
distributions entering recombination is coming from a jet. Eventually
we arrive at
\begin{multline}
  \frac{d (N_A Y_{AB}(\Delta\Phi))}{d P_{TA} d P_{TB}} = 
  2\pi I \, \hat v \, \frac{8 C_B M_{TB}}{P_{TB}} g_a (p_a) \\  \times 
  (2\pi \phi_0^2)^{-1/2} e^{-(\Delta\phi)^2/(2\phi^2_0)}
  I_0 \left( \frac{P_{TB} \sinh\eta_T}{2T} \right) 
  \\  \times K_1 \left( \frac{
  m^T_c \cosh\eta_T}{T}  \right) 
  D_{a/A} \left( \frac{P_{TA}}{P_{TA}+P_{TB}/2} \right)
\end{multline}
with $p_a = P_{TA} + P_{TB}/2 + \Delta E$.
The normalization constant $\hat v < 1$ arises from restricting the 
integration over $\Sigma$ to a subspace given by the jet cone.
Our notation implies that the fragmented hadron $A$ is the trigger hadron, 
but the process where $B$ is the trigger has to be taken into account as well.
Details of these calculations will be provided in a forthcoming publication.

\begin{figure}
  \epsfig{file=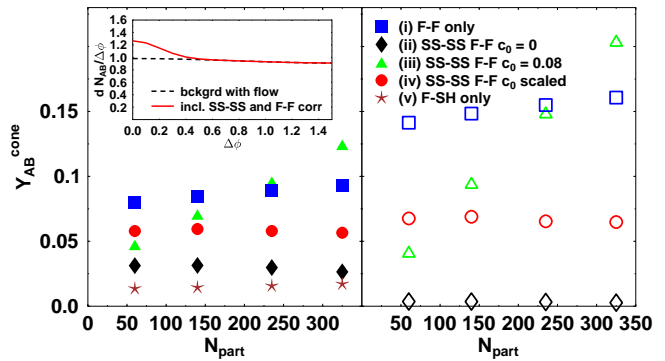,width=\columnwidth}
  \caption{\label{fig:results} $Y_{AB}^{\text{cone}}$ which is $Y_{AB}$ 
  integrated over $0\le \Delta\phi \le 0.94$, for meson (left panel) and 
  baryon triggers (right panel) as a 
  function of centrality. Insert: Associated yield as a function of 
  $\Delta \phi$ before background subtraction using SS-SS and F-F 
  contributions at an impact parameter $b=8$ fm.}
\end{figure}

Fig. \ref{fig:results} shows numerical examples of the different 
contributions. We use windows of 1.7 GeV/$c\le 
P_{TB} \le 2.5$ GeV/$c$ for associated particles and 2.5 GeV/$c\le 
P_{TA} \le 4.0$ GeV/$c$ for trigger particles and $|y|<0.35$ as in 
\cite{Sick:04}. Our calculation includes charged pions and kaons 
as well as protons and antiprotons. 
The parameters of the thermal parton
phase are taken from \cite{Fr03a}. 
We use the minijet distributions from 
\cite{FMS:02} and KKP fragmentation functions \cite{KKP:00}.
The adjustable parameters in our model are the azimuthal 
width $\phi_0$, which is chosen to be 0.2 for all processes, the 
correlation strength $\hat c_0$, and $\hat v$. 

Fig.\ \ref{fig:results} compares the background subtracted associated 
yield $Y_{AB}$, integrated over $0 \le \Delta \phi \le 0.94$ as a 
function of centrality for the following scenarios: (i) F-F process only, 
(ii) FF-FF and SS-SS with no soft correlations ($\hat c_0 = 0$), and 
(iii) the same with $\hat c_0 = 0.08$. The left panel is for meson 
triggers, the right panel for baryon triggers. The insert shows the 
uncorrelated background due to the associated yield from the SS-SS 
process and independent fragmentation as well as the signal 
for a near side correlation from SS-SS and F-F for $\hat c = 0.08$. 

As seen in the figure, the F-F mechanism (squares) produces strong 
near-side correlations, which are larger for baryon triggers. 
However, the trigger yields from this
process are small in the considered window, thus adding trigger 
particles from recombination
dilutes the signal dramatically (full diamonds).
This effect is even more pronounced for baryons (open diamonds). 
Switching on SS-SS correlations strongly increases the hadron correlations 
(triangles). Note that a constant value of $\hat c_0$ corresponds to a 
correlation volume $V_c$ scaling with $N_{\text{part}}$, which is not
likely realistic. We also show the result for a more realistic fixed
$V_c$ corresponding to $\hat c = 0.08 \times 100/N_{\text{part}}$ 
(circles), which makes $Y_{AB}$ vary weakly with centrality.

We also show F-SH correlations for $\pi-\pi$ and $\hat v = 0.5$ 
(stars). The yields in this case are so small that the contribution is 
negligible compared with SS-SS and F-F. It was already pointed out in 
\cite{Fr03a} that with the parton parametrization found by fitting the 
single inclusive spectra, soft-hard processes are subdominant. Other 
groups using different parametrizations have come to different conclusions 
\cite{Gr03a,Hw03a}.

The PHENIX preliminary results do not show clearly identifiable near-side 
jet cones above background for baryon triggers \cite{Sick:04}, which
suggests that one must be cautious with the interpretation of the 
$\Delta \phi$-integrated associated yields in this case. In our model the 
relative behavior of baryon and meson triggers depends on the relative 
strength of the SS-SS and F-F contributions. Double fragmentation (F-F)
would predict a larger associated yield for baryons triggers than for
meson triggers. For the most realistic case (F-F and SS-SS with $\hat c 
\approx 0.08$ and fixed correlation volume) meson and baryon triggers result 
in dihadron 
correlations of approximately equal magnitude. More experimental information
is needed, including measurements of $p-p$ collisions, to determine which
scenario describes the data best.

In summary, we have shown that correlations among partons in a quark-gluon
plasma naturally translate into correlations between hadrons formed by
recombination of quarks with an amplification factor similar to the one
obtained for elliptic flow. Preliminary data from the PHENIX collaboration 
are consistent with hadron production by quark recombination if two-parton 
correlations of order 10\% within a fixed correlation volume are assumed
to be present in the deconfined phase. We conclude that the existence of 
localized angular correlations among hadrons are not in contradiction to 
the recombination scenario, since a plausible mechanism for the formation
of such parton correlations --- passage of hard partons through the dense
matter --- exists in nuclear collisions at RHIC.

\begin{acknowledgments}
The authors want to thank B.\ Jacak, R.\ Lacey and A.\ Sickles for 
illuminating discussions. This work was supported in part by RIKEN, 
Brookhaven National Laboratory and DOE grants DE-AC02-98CH10886, 
DE-FG02-96ER40945 and DE-FG02-03ER41239. 
\end{acknowledgments}

\end{document}